\documentclass[superscriptaddress,amsmath,amssymb,aps,reprint,prb,twocolumn]{revtex4-2}
\usepackage[percent]{overpic}
\usepackage{dcolumn}
\usepackage{bm}
\usepackage{hyperref}
\usepackage{svg}
\usepackage{graphicx}
\usepackage[normalem]{ulem}
\hypersetup{
	citecolor = blue,
	colorlinks = true,
	urlcolor = blue
}
\usepackage{comment}

\usepackage{tikz}
\usepackage{mathrsfs}

\usepackage{braket}

\renewcommand{\Re}{\operatorname{Re}}
\renewcommand{\Im}{\operatorname{Im}}

\begin{document}

\title{\textbf{Stability-threshold control of helicity-selective antiferromagnetic resonance}}

\author{Masato Todani}
\affiliation{Institute for Solid State Physics, University of Tokyo, Kashiwa 277-8581, Japan}

\author{Satoshi Iihama}
\affiliation{Department of Materials Physics, Nagoya University, Nagoya 464-8603, Japan}

\author{Yuto Moritake}
\affiliation{Institute of Industrial Science, The University of Tokyo, 4-6-1 Komaba, Meguro-ku, Tokyo 153-8505, Japan}

\author{Takeo Kato}
\affiliation{Institute for Solid State Physics, University of Tokyo, Kashiwa 277-8581, Japan}

\date{\today}

\begin{abstract}
We study helicity-selective antiferromagnetic resonance in an antiferromagnetic-insulator/nonmagnetic-metal junction with sublattice-dependent damping and spin-orbit torque. 
By formulating the linearized Landau--Lifshitz--Gilbert equation as a \(2\times2\) non-Hermitian eigenvalue problem, we analyze the complex resonance frequencies and the stability threshold, using MnF$_2$ parameters as a representative example of a uniaxial antiferromagnet.
We show that the absorption is strongly enhanced near the stability threshold, where one helicity mode becomes weakly damped, leading to linewidth narrowing and pronounced helicity selectivity in the sub-THz regime. 
The selected helicity can be switched by reversing the current direction, which reverses the spin-orbit torque acting on the interfacial sublattice. 
These results identify stability-threshold control as a design principle for electrically tunable helicity-selective antiferromagnetic resonance in uniaxial antiferromagnets.
\end{abstract}

\maketitle

\section{Introduction}
\label{sec:introduction}

Non-Hermitian physics provides a useful framework for describing open systems with dissipation and gain \cite{Ashida2020NonHermitianPhysics,Miri2019ExceptionalPointsOpticsPhotonics}. 
In magnetic systems, Gilbert damping, spin pumping, and current-induced torques provide experimentally accessible sources of non-Hermiticity, making magnetic resonance a natural platform for non-Hermitian dynamics \cite{Hurst2022NonHermitianPhysicsMagneticSystems,Yu2024NonHermitianTopologicalMagnonics}. 
Previous studies have often focused on exceptional points (EPs), where eigenvalues and eigenvectors coalesce, and on the associated response enhancement and mode selectivity in ferromagnets and synthetic antiferromagnets \cite{Heiss2012PhysicsExceptionalPoints,Lee2015,Galda2016,Liu2019,Wang2020,Duine2023,Wang2024}.

Intrinsic antiferromagnets offer a qualitatively different setting because their resonance frequencies are governed primarily by the exchange interaction and typically lie in the sub-THz/THz range \cite{Baltz2018AntiferromagneticSpintronics,Wang2018NiOAFMR}. This high-frequency regime makes antiferromagnets attractive for ultrafast spintronic and THz applications, but also implies that design principles established for ferromagnets and synthetic antiferromagnets do not apply directly to intrinsic antiferromagnets.
It is therefore important to clarify how non-Hermiticity affects antiferromagnetic resonance and how it appears in experimentally observable helicity-resolved responses.

Antiferromagnetic insulator (AFI)/nonmagnetic metal (NM) junctions provide a natural platform for introducing non-Hermitian effects into antiferromagnetic dynamics in the THz regime through 
sublattice-dependent interfacial damping and current-induced torques~\cite{Tserkovnyak2002EnhancedGilbertDamping,Tserkovnyak2005NonlocalMagnetizationDynamics,Cheng2014,Hals2011CurrentInducedDynamicsAntiferromagnets,Zelezny2016SOTAFM,Manchon2019CurrentInducedSpinOrbitTorques,Moriyama2018,Kamra2018}. 
Previous studies of such junctions have explored threshold instability, spin pumping, and current-driven auto-oscillation in the sub-THz/THz regime~\cite{Cheng2016,Khymyn2017AntiferromagneticTHzFrequencyJosephsonLikeOscillator,Guo2022NiOSpinPumping}. 
In the present work, we focus on the stable side of the threshold and show that the helicity-selective absorption is controlled primarily by the effective damping rates of the resonance modes. 
When the spin-orbit torque acts as antidamping, one helicity mode can become weakly damped near the stability threshold, producing strong absorption enhancement and linewidth narrowing. 
This mechanism is distinct from EP-induced response enhancement, because the enhanced response is governed by proximity to the stability threshold rather than by eigenmode coalescence.
This threshold-based mechanism exploits the helicity structure of intrinsic antiferromagnetic resonance, distinguishing it from non-Hermitian mode selection in ferromagnets and synthetic antiferromagnets.

\begin{figure}[tb]
  \centering
    \includegraphics[width=0.85\linewidth]{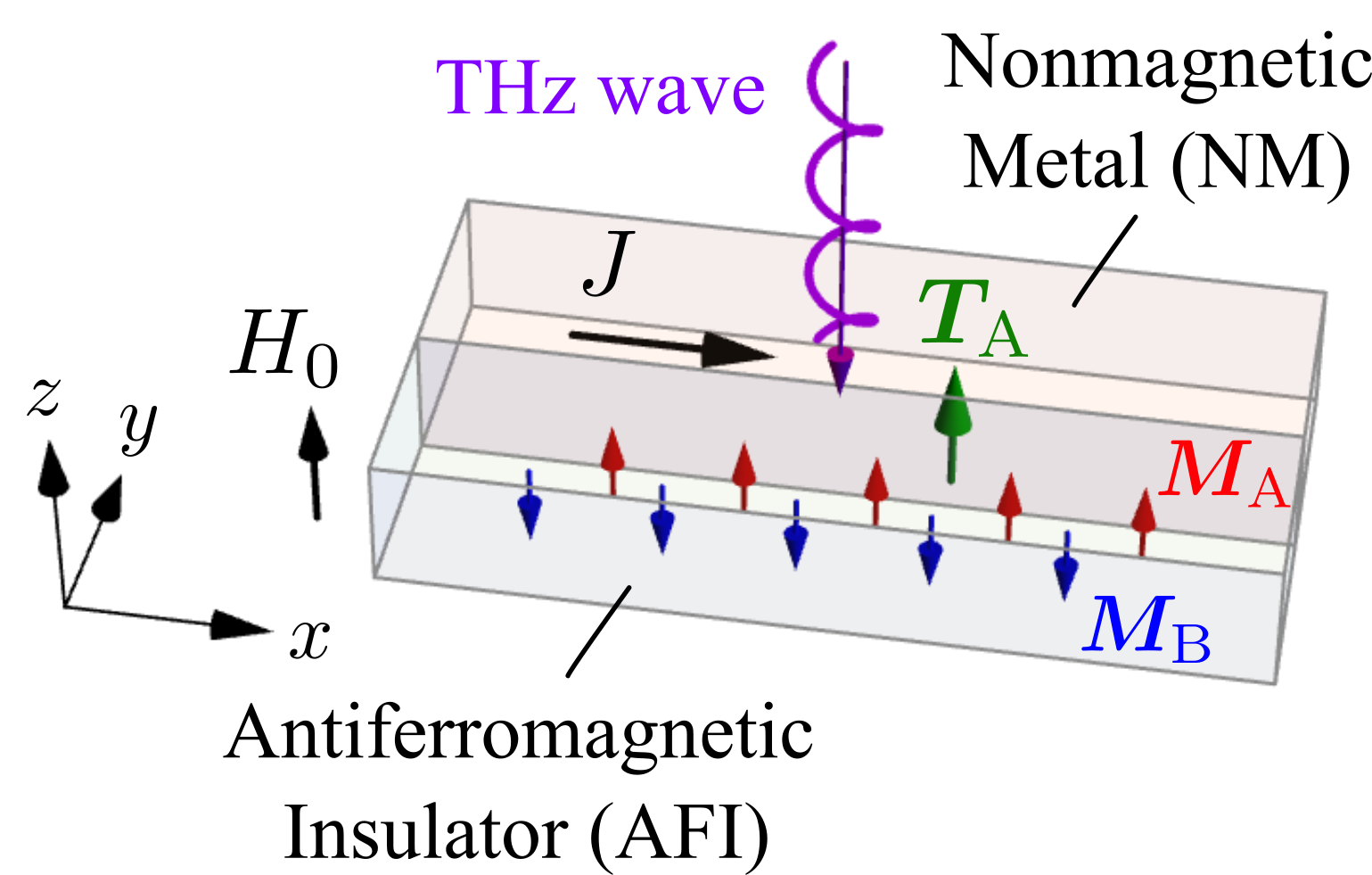}
  \caption{Schematic illustration of the antiferromagnetic insulator (AFI)/nonmagnetic metal (NM) junction considered in this study.
The antiferromagnet consists of two sublattices, $A$ and $B$, and is assumed to couple to the NM only through sublattice $A$.
A charge current $J$ flowing in the NM along the $x$ direction induces a damping-like spin-orbit torque ${\bm T}_{\rm A}$ acting on the interfacial sublattice \(A\).
A static magnetic field $H_0$ is applied along the $z$ direction.
The antiferromagnetic resonance driven by a circularly polarized THz wave is studied.}
  \label{fig:system_picture}
\end{figure}

In this work, we analyze helicity-selective antiferromagnetic resonance in a two-sublattice uniaxial antiferromagnet with sublattice-dependent Gilbert damping and a damping-like spin-orbit torque acting on the interfacial sublattice. 
We formulate the linearized Landau--Lifshitz--Gilbert equation as a \(2\times2\) non-Hermitian eigenvalue problem and derive the complex resonance frequencies and the stability-threshold condition.
Using MnF$_2$ parameters, we calculate the response to a circularly polarized ac magnetic field and show that the absorption near the stability threshold exhibits pronounced helicity selectivity that can be switched by reversing the current direction. 
We also analyze the EP condition to clarify why the enhanced response in the present parameter regime is governed by the stability threshold rather than by the EP.
The key point is that the same current-induced non-Hermiticity controls both damping compensation and helicity selection, enabling a switchable circular-polarization response without requiring proximity to an EP.

\section{Formulation}
\label{sec:formulation}

\subsection{Setup}

We consider a two-sublattice antiferromagnetic insulator (AFI) described by the Hamiltonian
\begin{align}
\mathcal{H}_{\rm AF} =
J_{\rm AF} \sum_{\langle i,j\rangle}
\bm{S}_i\cdot\bm{S}_j - D \sum_i (S_i^z)^2
- \gamma\hbar \sum_i \bm{H}\cdot\bm{S}_i .
\label{eq:Ham}
\end{align}
Here, \(J_{\rm AF} > 0 \) denotes the antiferromagnetic exchange interaction, \(D\) is the easy-axis anisotropy, and \(\gamma\) is the gyromagnetic ratio. 
In the following, \(\bm M_A\) and \(\bm M_B\) denote the coarse-grained sublattice magnetizations associated with the localized spins in Eq.~\eqref{eq:Ham}, with magnitude \(M_s\).
The external magnetic field is given by
\begin{align}
\bm H &= H_0\hat{\bm z}+\bm H_{\rm ac}(t),
\label{eq:Hact}
\end{align}
where \(H_0\) is the static magnetic field, \(\bm H_{\rm ac}(t)= (h\cos\omega t,-h\sin\omega t,0)\) is the in-plane ac magnetic field, and $\omega$ denotes the driving frequency.
With this convention, positive and negative $\omega$ correspond respectively to
clockwise and counterclockwise rotations of the ac field in the $xy$ plane, as viewed from the $+z$ direction.
We assume a bipartite AFI with equilibrium magnetizations aligned parallel and antiparallel to the $z$ axis on sublattices \(A\) and \(B\), respectively.

We consider a junction between the AFI and a nonmagnetic metal (NM), as shown in Fig.~\ref{fig:system_picture}.
We assume that only the \(A\) sublattice is coupled to the NM.
For MnF$_2$, a representative two-sublattice uniaxial AFI, the \((100)\) surface is a natural candidate for realizing sublattice-selective interfacial coupling.
In this geometry, interfacial electronic states are expected to couple more strongly to one antiferromagnetic sublattice than to the other.
Such sublattice-selective coupling provides the simplest mechanism for generating unequal interfacial damping and a spin-orbit torque acting predominantly on one sublattice.
Although realistic interfaces may involve roughness, reconstruction, or mixed termination, the present model captures the simplest limit in which the non-Hermitian asymmetry originates from sublattice-selective interfacial coupling.

\subsection{LLG equations and linear response}

The Landau--Lifshitz--Gilbert (LLG) equations for the two sublattices are given by
\begin{align}
\frac{d\bm M_A}{dt}
&=
\gamma \bm H_{{\rm eff},A}\times \bm M_A
+
\frac{\alpha_1}{M_s}
\bm M_A\times \frac{d\bm M_A}{dt} \notag \\
& \qquad 
+ \frac{\gamma c_J}{M_s}
\bm M_A\times
(\bm T_A\times \bm M_A),
\\
\frac{d\bm M_B}{dt}
&=
\gamma \bm H_{{\rm eff},B}\times \bm M_B
+
\frac{\alpha_2}{M_s}
\bm M_B\times \frac{d\bm M_B}{dt}.
\end{align}
Here, \(c_J\) denotes the effective field characterizing the damping-like spin-orbit torque acting on the interfacial \(A\) sublattice. 
The parameter \(c_J\) is controlled by the magnitude and direction of the current in the NM. 
Depending on its sign, this term acts as either effective damping or effective antidamping. The spin polarization direction of the itinerant electrons is assumed to be
\begin{align}
\bm T_A=\hat{\bm z}.
\end{align}
The effective magnetic fields are given by
\begin{align}
\bm H_{{\rm eff},A} & = \bm H -H_J \frac{\bm M_B}{M_s} + H_D \frac{M_{A,z}}{M_s} \hat{\bm z}, \\
\bm H_{{\rm eff},B} &= \bm H - H_J \frac{\bm M_A}{M_s} + H_D \frac{M_{B,z}}{M_s} \hat{\bm z},
\end{align}
where 
\begin{align}
H_J& =\frac{S z J_{\text{AF}}}{\gamma\hbar}, \\
H_D&= \frac{2DS}{\gamma\hbar} .
\end{align}
Here, \( z \) is the coordination number.

The equilibrium magnetizations of the two sublattices are assumed to be
\begin{align}
\bm{M}_A^{(0)}=M_s\hat{\bm z}, \qquad \bm{M}_B^{(0)}=-M_s\hat{\bm z}.
\end{align}
This collinear state remains a stationary solution even in the presence of a finite \(c_J\) because the damping-like torque vanishes for \(\bm M_A \parallel \bm T_A \parallel \hat{\bm z}\).
We introduce a small transverse modulation of the magnetization around the equilibrium state as
\begin{align}
\bm M_\nu = \bm M_\nu^{(0)} + \delta\bm M_\nu(t).
\label{magnon_approximation}
\end{align}
We linearize the LLG equation with respect to the magnetization modulation, assuming $|\delta M_\nu^{x,y}(t)|\ll M_s$ 
($\nu={\rm A},{\rm B}$).
To simplify the calculation, we introduce a complex representation of the magnetic field
\begin{align}
h^+(t)=h_x+ih_y=he^{-i\omega t},
\end{align}
and a complex deviation of the magnetization
\begin{align}
\delta M_\nu^+ = \delta M_\nu^x+i\delta M_\nu^y =
\tilde M_\nu(\omega)e^{-i\omega t},
\end{align}
where $\tilde M_\nu(\omega)$ denotes the oscillation amplitude on sublattice $\nu$.
The linearized LLG equation for the amplitudes is written as
\begin{align}
\begin{pmatrix}
\mathcal A_A(\omega) & -\Omega\\
\Omega & \mathcal A_B(\omega)
\end{pmatrix}
\begin{pmatrix}
\tilde M_A(\omega)\\
\tilde M_B(\omega)
\end{pmatrix}
= - \gamma M_s h \begin{pmatrix}
1 \\
-1
\end{pmatrix},
\label{eq:LLGM}
\end{align}
where 
\begin{align}
\mathcal A_A(\omega) &= -(1-i\alpha_1)\omega - \gamma(H_0+H_J+H_D+ic_J), \\
\mathcal A_B(\omega) &= -(1+i\alpha_2)\omega - \gamma(H_0-H_J-H_D), \\
\Omega &= \gamma H_J .
\end{align}
The $2\times 2$ matrix in Eq.~\eqref{eq:LLGM} is non-Hermitian due to the sublattice-dependent Gilbert damping constants and the SOT.
Therefore, the magnetization dynamics is described within the framework of non-Hermitian physics.
In particular, \(c_J\) acts as a non-Hermitian control parameter for the interfacial \(A\) sublattice.
Depending on its sign and magnitude, it acts as either effective damping or antidamping, and can eventually drive the system into the gain regime.

Solving Eq.~\eqref{eq:LLGM}, we obtain
\begin{align}
\begin{pmatrix}
\tilde M_A(\omega)\\
\tilde M_B(\omega)
\end{pmatrix}
= - \frac{\gamma M_s h}{F(\omega)}
\begin{pmatrix}
\mathcal A_B(\omega) & \Omega\\
-\Omega & \mathcal A_A(\omega)
\end{pmatrix}
\begin{pmatrix}
1 \\
-1
\end{pmatrix},
\label{eq:M_solution}
\end{align}
where 
\begin{align}
F(\omega) &= \mathcal A_A(\omega)\mathcal A_B(\omega)+\Omega^2 ,
\label{eq:Fomega}
\end{align}
is the characteristic function.

\subsection{Resonance frequencies and stability thresholds}

The resonance frequencies are determined by the condition
$F(\omega)=0$, which leads to a quadratic equation $a\omega^2+b\omega+c=0$ with
\begin{align}
a &= (1-i\alpha_1)(1+i\alpha_2), \\
b &= \gamma \Bigl[ (1-i\alpha_1)(H_0-H_J-H_D) \nonumber\\
&\qquad + (1+i\alpha_2)(H_0+H_J+H_D+i c_J) \Bigr], \\
c &= \gamma^2 \Bigl[ (H_0+H_J+H_D+i c_J)(H_0-H_J-H_D) \nonumber \\
& \qquad + H_J^2 \Bigr].
\end{align}
The two resonance frequencies are given by
\begin{align}
\omega_\pm =
\frac{-b\pm\sqrt{b^2-4ac}}{2a}.
\label{eq:omega_pm}
\end{align}
Modes with $\Im \, \omega<0$ decay exponentially with time, whereas modes with $\Im\,\omega>0$ grow exponentially, indicating the onset of dynamical instability.
The linear stability of the system is characterized by
$\Gamma_{\max} = \max [ \, \Im\, \omega_+,\Im\, \omega_- ]$ as
\begin{align}
\Gamma_{\max}<0 \ : \ &\text{stable}, \\
\Gamma_{\max}>0 \ : \ &\text{unstable}.
\end{align}
Therefore, the stability threshold is given by $\Gamma_{\max}=0$.

The exceptional point (EP), at which the two eigenvalues and eigenvectors coalesce, is defined by the discriminant condition \(\mathcal D=b^2-4ac=0\) (see also Appendix~\ref{app:EP}). 
As shown below, however, the enhanced helicity-selective response in the present parameter regime is governed by the stability threshold rather than by the EP.

\subsection{Absorption spectrum}

Next, we calculate the sub-THz absorption rate as a function of the real driving frequency $\omega$. 
The absorption power is defined as the time-averaged work done by the ac magnetic field on the magnetization:
\begin{align}
P(\omega) = \sum_{\nu = A, B} \left\langle \bm H_{\rm ac}(t)\cdot \partial_t\bm M_\nu(t) \right\rangle .
\end{align}
By substituting Eq.~\eqref{eq:M_solution}, we obtain
\begin{align}
P(\omega) &= \gamma M_s h^2\omega\, \Im \left[ \frac{2\Omega+\mathcal A_A(\omega)-\mathcal A_B(\omega)}{F(\omega)} \right].
\label{eq:absorption_power}
\end{align}
For convenience, we normalize the absorption rate as
$\tilde{P}(\omega)=P(\omega)/P_0$, where $P_0 = \gamma M_s h^2/2$.

\section{Results}
\label{sec:results}

\subsection{Parameter settings}

In the numerical calculations, we employ parameters for \text{MnF}$_\text{2}$, a representative antiferromagnetic insulator ~\cite{Vaidya2020}.
The temperature is assumed to be well below the N\'eel temperature.
Unless otherwise specified, we set \(\gamma = 1.76\times 10^{11}\, {\rm rad\,s^{-1}T^{-1}}\), \(H_J = 47.05\, {\rm T}\), and \(H_D = 0.82\, {\rm T}\).
To estimate the order of magnitude of the SOT effective field, we treat \(c_J\) as a phenomenological parameter and take \(|c_J|\le 0.08\,{\rm T}\), comparable to effective fields discussed in spin-orbit-torque-driven magnetic systems~\cite{Collet2016,Wang2020}.
Our aim is to clarify the basic mechanisms of threshold control and circular-polarization selectivity rather than to make a quantitative comparison with a specific material platform.

\begin{figure}[t]
  \centering
  \includegraphics[width=1.0\linewidth]{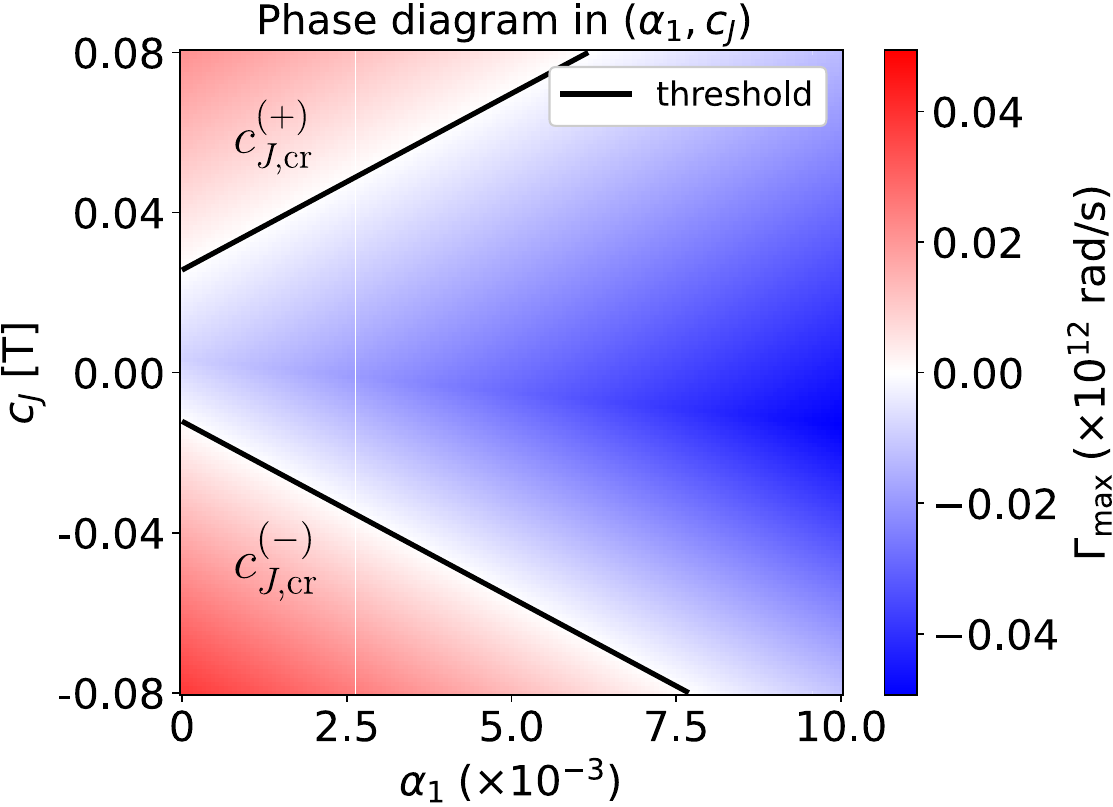}
  \caption{
  Stability diagram in the \((\alpha_1,c_J)\) plane for \(H_{0} = 0 \) and \(\alpha_2 = 0.002\).
  The color map shows \(\Gamma_{\max}=\max\, [\Im\omega_+,\Im\omega_-]\).
  The black line indicate the stability thresholds \(\Gamma_{\max}=0\).
  }
  \label{fig:phase_diagram}
\end{figure}

We first investigate the stability of the system.
Figure~\ref{fig:phase_diagram} shows a color map of \(\Gamma_{\max}\) in the \((\alpha_1,c_J)\) plane for \(H_{0} = 0\, {\rm T}\) and \(\alpha_2 = 0.002\).
The blue (red) region corresponds to the stable (unstable) region with $\Gamma_{\max}<0$ ($\Gamma_{\max}>0$).
The instability appears only when the spin-orbit torque is sufficiently large.
The black line indicates the stability boundary \( \Gamma_{\max}=0 \).
For the MnF$_2$ parameters considered here, the EP lies far outside the range shown in Fig.~\ref{fig:phase_diagram}; its location and damping rate are discussed in Appendix~\ref{app:EP}.

The approximately linear stability boundary seen in Fig.~\ref{fig:phase_diagram} is consistent with the analytical expression for the threshold line.
For $H_D,H_0,|c_J|\ll H_J$ and $\alpha_1,\alpha_2\ll 1$, the critical value of $c_J$ for the stability boundary determined by $\Gamma_{\rm max} = 0$ is given by
\begin{align}
c_{J,{\rm cr}}^{(\lambda)} &= - \frac{(\lambda R-H_0) }{\lambda R-H_D-H_J} \Bigl[ (H_D+H_J)(\alpha_1+\alpha_2) \nonumber \\
& \hspace{15mm} +\lambda R (\alpha_2-\alpha_1) \Bigr] ,
\end{align}
where $\lambda=\pm 1$ denotes the helicity of the unstable mode at the stability boundary, and $R = \sqrt{H_D(H_D+2H_J)}$ (for a detailed derivation, see Appendix~\ref{app:analytic}).
This expression indicates that the stability boundary is linear since $c_{J,{\rm cr}}^{(\lambda)}$ depends linearly on $\alpha_1$ and $\alpha_2$.

\subsection{Complex resonance frequencies}

\begin{figure}[t]
  \includegraphics[width=1.0\linewidth]{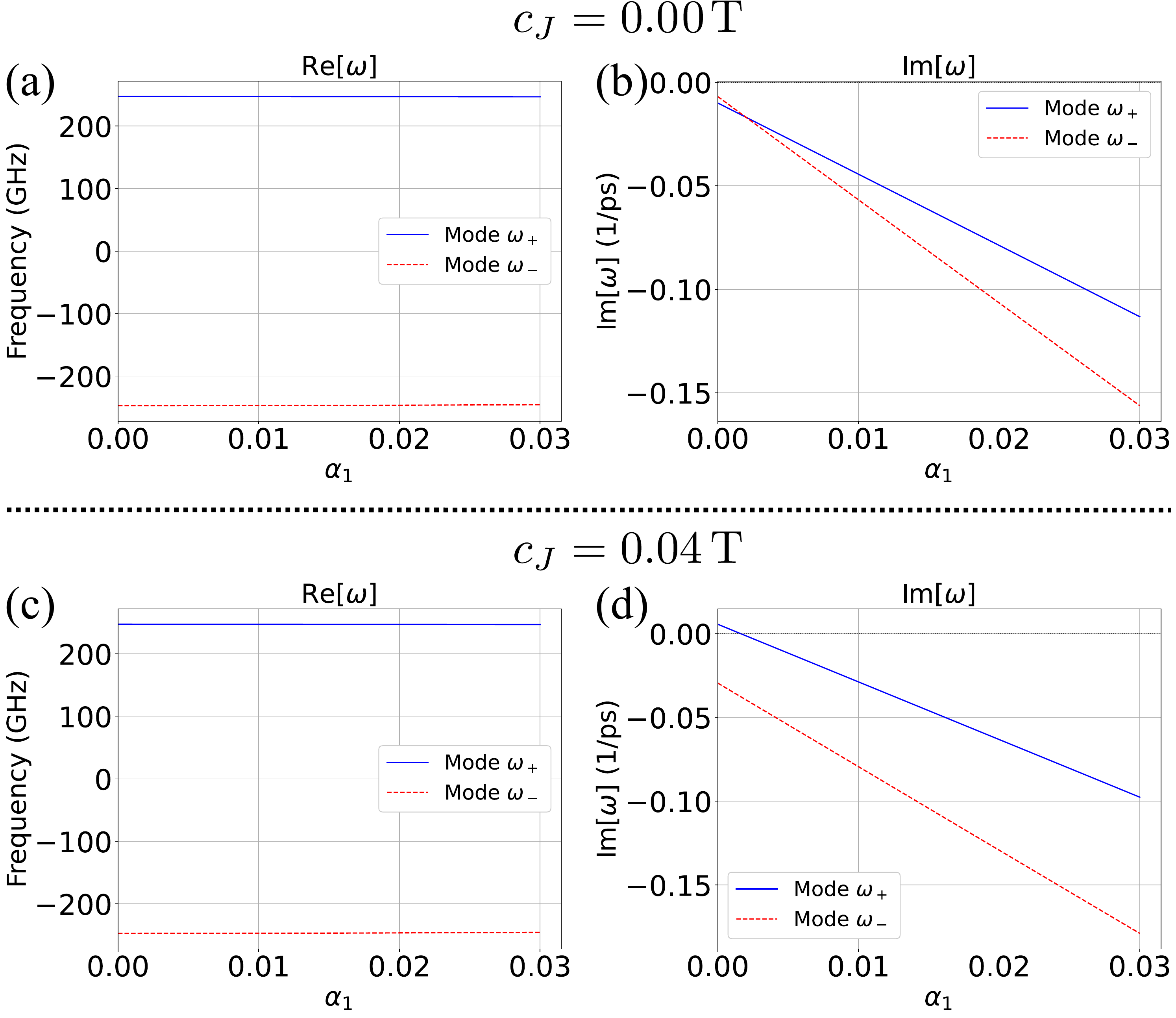}
  \caption{
  Real and imaginary parts of the two resonance frequencies as a function of \(\alpha_1\) for (a,b) \(c_J = 0.00\, {\rm T}\) and (c,d) \(c_J = 0.04\, {\rm T}\). 
  The remaining parameters are set to \(H_0 = 0\) and \(\alpha_2 = 0.002\).}
  \label{fig:eigenfrequency}
\end{figure}

Figures~\ref{fig:eigenfrequency}(a) and (b) show \(\Re\, \omega_\pm\) and \(\Im \, \omega_\pm\) as a function of \(\alpha_1\) for $c_J = 0$.
The real parts \(\Re \omega_\pm\) represent the resonance frequencies, while the imaginary parts \(\Im\, \omega_\pm\) represent the decay rate for \(\Im\omega_\pm<0\) and the growth rate for \(\Im\omega_\pm>0\).
The sign of \(\Re\omega_\pm\) determines which helicity of the circularly polarized driving field is resonant.
As shown in Fig.~\ref{fig:eigenfrequency}(a) and (b), the real parts \(\Re \omega_\pm\) are almost independent of $\alpha_1$ in the plot range, while the imaginary parts \(\Im\, \omega_\pm\) depend linearly on $\alpha_1$.
Although the difference between the imaginary parts of the two modes gives rise to helicity-dependent absorption, this effect is weak because the magnitudes of \(\Im\,\omega_\pm\) remain comparable.
Figures~\ref{fig:eigenfrequency}(c) and (d) respectively show \(\Re \, \omega_\pm \) and \(\Im\, \omega_\pm \) for \(c_J = 0.04 \, {\rm T}\).
In contrast to the case of  \(c_J = 0 \), 
\(\Im\, \omega_+\) changes sign at $\alpha_1 = 0.0016$, which corresponds to the stability boundary, whereas \(\Im\, \omega_-\) remains negative.
When the parameters are chosen near this sign change on the stable side, the damping rates of the two helicity modes become substantially different, leading to strong helicity-dependent absorption, as discussed in the next subsection.

\subsection{Absorption spectra}

\begin{figure}[t]
  \centering
  \includegraphics[width=0.8\linewidth]{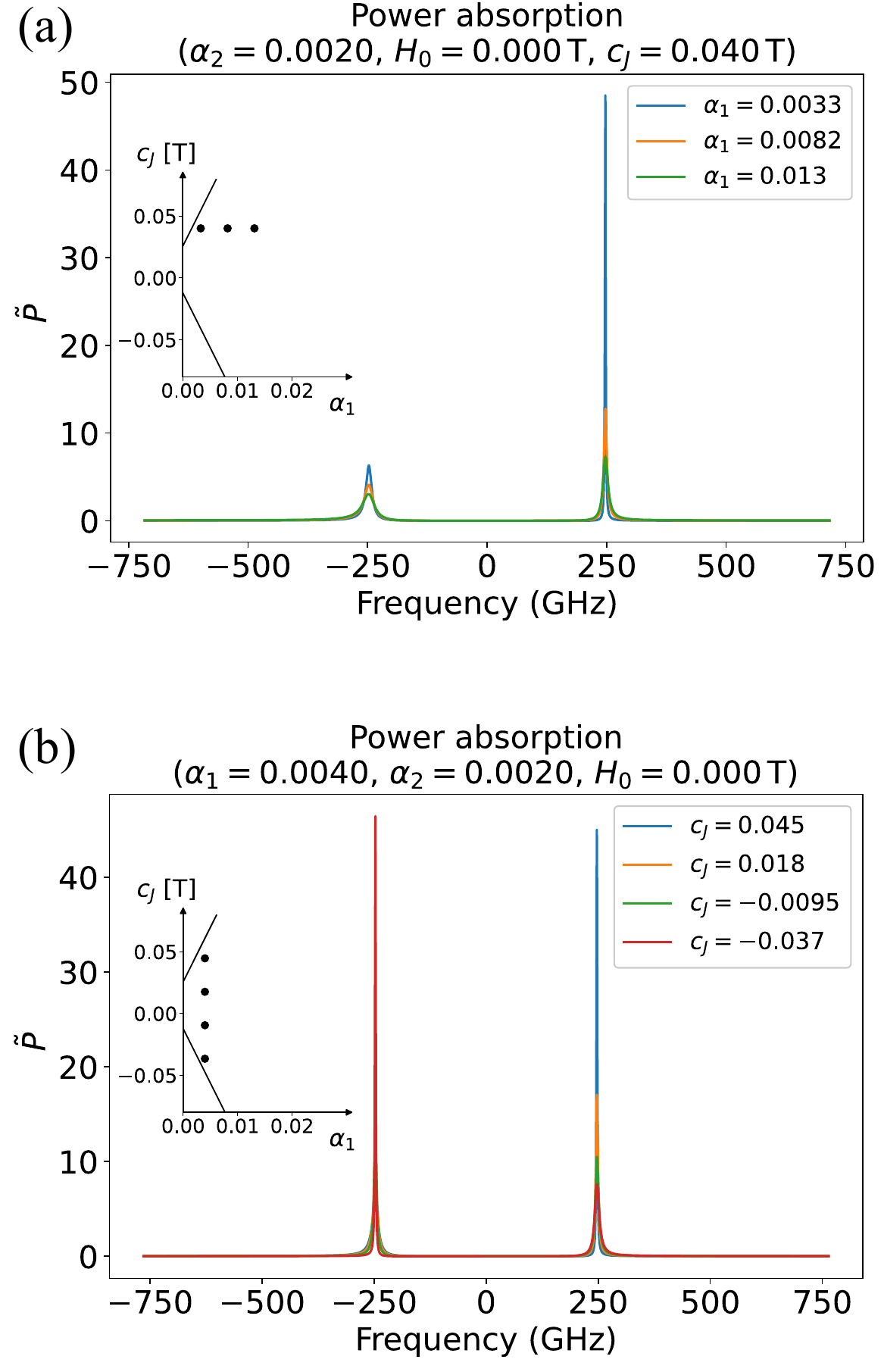}
  \caption{Frequency dependence of the normalized absorption rate (a) for several values of $\alpha_1$ at fixed spin-orbit torque $c_J$ and (b) for several values of $c_J$ at fixed damping constant $\alpha_1$. 
  The insets indicate the corresponding parameter sets in the \((\alpha_1,c_J)\) diagram.}
  \label{fig:absorption_spectrum}
\end{figure}

Figure~\ref{fig:absorption_spectrum}(a) shows the absorption spectra for \(c_J = 0.04\, {\rm T}\) as \(\alpha_1\) approaches the stability threshold from the stable side ($\alpha_1 > 0.0016$).
Far from the stability threshold, the absorption peaks have broad linewidths and relatively small peak intensities.
In contrast, as the system approaches the stability boundary from the stable side, one peak becomes narrower and stronger, reflecting the approach of \(\Im \, \omega_+\) to zero.
This peak narrowing occurs only on the positive-frequency branch, corresponding to one helicity of the circularly polarized sub-THz ac magnetic field.
This demonstrates that the absorption enhancement is helicity selective.

Figure~\ref{fig:absorption_spectrum}(b) shows the absorption spectra for $\alpha_1 = 0.004$ in the stable regime for several values of the damping-like spin-orbit torque, $c_J=-0.0366\, {\rm T}, -0.0095\, {\rm T}, 0.0176\, {\rm T}$, and $0.0447\, {\rm T}$ (see also the inset). By reversing the sign of \(c_J\), the helicity associated with the stronger resonant peak is reversed. Since \(c_J\) is controlled by the charge current in the NM, the helicity exhibiting the stronger resonant absorption can be switched simply by reversing the current direction. This behavior demonstrates that the spin-orbit torque controls not only the absorption enhancement near the stability threshold but also the helicity selectivity. Therefore, tuning \(c_J\) provides electrical control of both the absorption enhancement and the circular-polarization selectivity.

These results demonstrate that the helicity selectivity is not merely a consequence of the equilibrium mode frequencies, but is strongly amplified by non-Hermitian damping compensation near the stability threshold.
Thus, the present model provides a proof-of-principle mechanism for electrically tunable polarization-selective responses based on threshold-controlled antiferromagnetic resonance.

\subsection{Circular polarization of the transmitted wave}

To characterize the experimentally relevant polarization conversion for a linearly polarized incident wave, we consider the degree of circular polarization of the transmitted wave. Since a linearly polarized wave can be decomposed into an equal-weight superposition of the two opposite helicities, we introduce a simple phenomenological model for the transmitted intensities by assuming Beer--Lambert-type attenuation. 
In this model, the attenuation factor is assumed to be exponential in the absorption $P(\omega)$, with the proportionality constant determined by the AFI thickness.
The transmitted intensities are then written as
\begin{align}
I^{\rm out}(\omega)&=\frac{I_{\rm in}}{2}\exp\left[-\frac{P(\omega)}{P_{\rm ref}}\right],\\
I^{\rm out}(-\omega)&=\frac{I_{\rm in}}{2}\exp\left[-\frac{P(-\omega)}{P_{\rm ref}}\right],
\end{align}
where \(I_{\rm in}\) denotes the incident intensity of the linearly polarized wave and \(P_{\rm ref}\) is treated as an adjustable scale parameter inversely proportional to the AFI thickness. We then define the degree of circular polarization of the transmitted wave as
\begin{align}
\mathcal C(\omega)
=
\frac{I^{\rm out}(\omega)-I^{\rm out}(-\omega)}
     {I^{\rm out}(\omega)+I^{\rm out}(-\omega)}.
\end{align}
Substituting the transmitted intensities, we obtain
\begin{align}
\mathcal C(\omega)
=
\frac{
\exp[-P(\omega)/P_{\rm ref}]-\exp[-P(-\omega)/P_{\rm ref}]
}{
\exp[-P(\omega)/P_{\rm ref}]+\exp[-P(-\omega)/P_{\rm ref}]
}.
\end{align}
In this definition, \(\mathcal C>0\) and \(\mathcal C<0\) correspond to the dominance of the \(+\omega\) and \(-\omega\) helicity components, respectively.

\begin{figure}[t]
  \centering
  \includegraphics[width=1.0\linewidth]{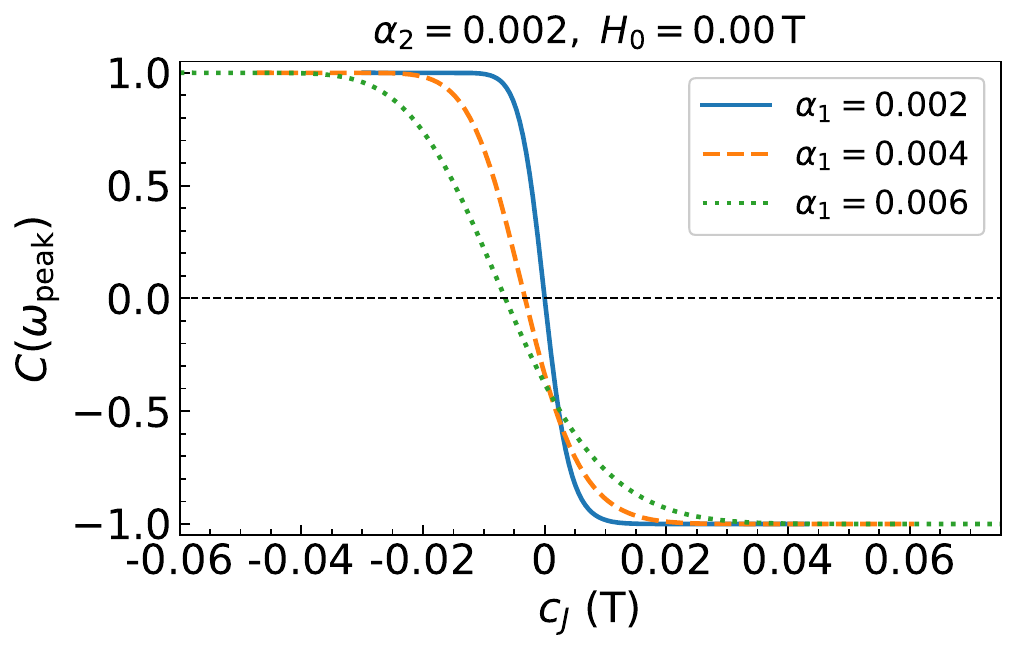}
  \caption{
    Degree of circular polarization \(\mathcal C(\omega_{\rm peak})\) as a function of \(c_J\) for \(\alpha_1=0.002,\,0.004,\) and \(0.006\), where \(\omega_{\rm peak}=(|\operatorname{Re}\omega_+|+|\operatorname{Re}\omega_-|)/2\). The other parameters are \(\alpha_2=0.002\), \(H_0=0\), and \(P_{\rm ref}=2P_0\). Only the linearly stable region is shown.
    }
  \label{fig:eta_map}
\end{figure}

Figure~\ref{fig:eta_map} shows \(\mathcal C(\omega_{\rm peak})\) as a function of \(c_J\) for \(\alpha_1=0.002,\,0.004,\) and \(0.006\).
Here, \(\omega_{\rm peak}\) is defined as \((|\operatorname{Re}\omega_+|+|\operatorname{Re}\omega_-|)/2\), because the two resonance frequencies have nearly the same magnitude.
The other parameters are set to \(\alpha_2=0.002\), \(H_0=0\), and \(P_{\rm ref}=2P_0\).
The sign of \(\mathcal C\) reverses between positive and negative values of \(c_J\).
This means that the preferred helicity of the transmitted wave can be switched simply by reversing the current direction in the NM.
This behavior originates from the fact that the sign of \(c_J\) determines which helicity becomes long-lived near the stability threshold, while the opposite helicity remains strongly damped.
Therefore, the present result supports stability-threshold control as a design principle for polarization-selective sub-THz responses in antiferromagnets.

\subsection{Self-oscillation in the unstable region}

\begin{figure}[tb]
  \centering
  \includegraphics[width=1.00\linewidth]{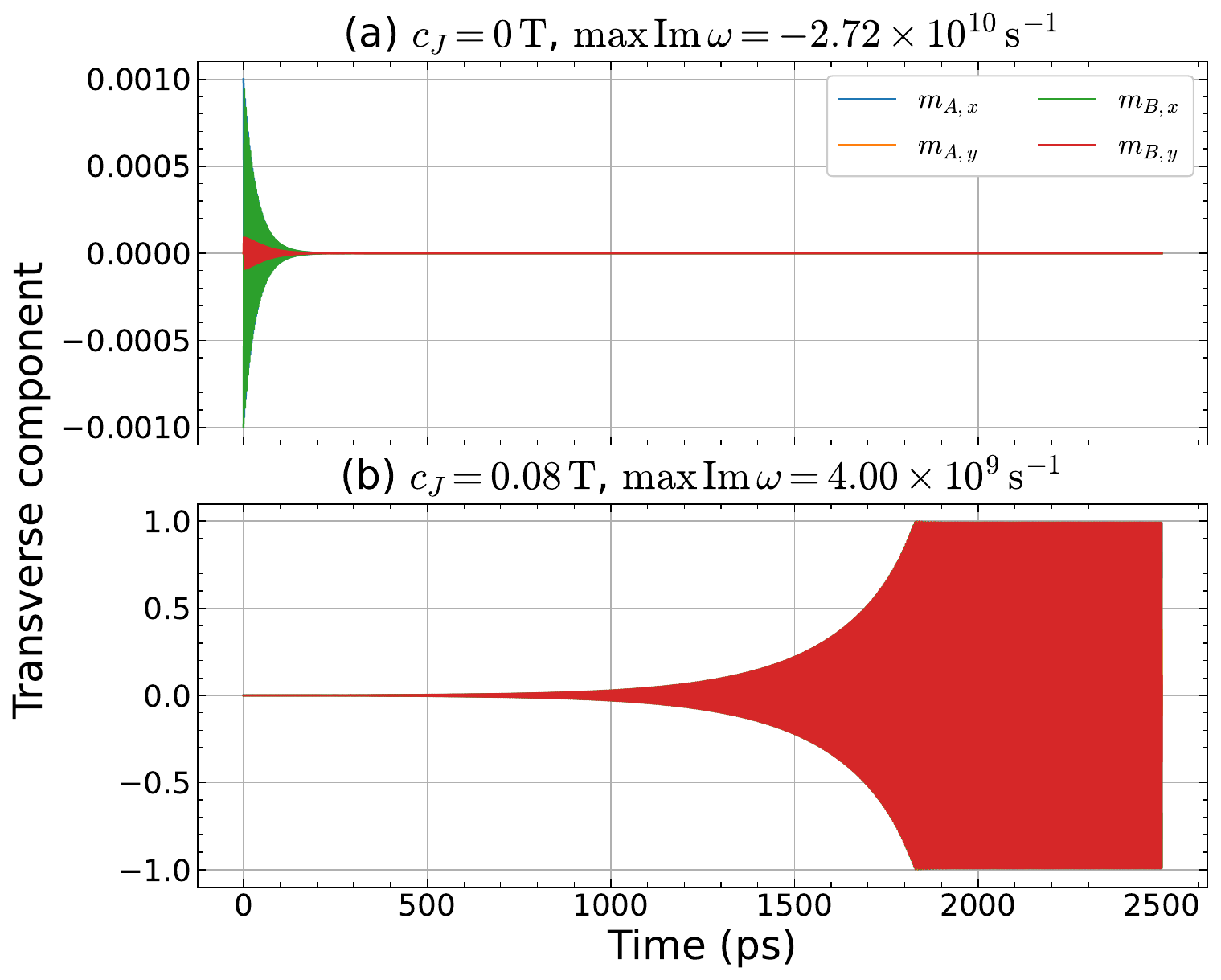}
  \caption{
    Real-time magnetization dynamics for (a) $c_J=0$ (the stable region), and (b) $c_J = 0.08\,{\rm T}$ (deeper in the unstable region) obtained by solving the full LLG equation.
    The remaining parameters are set to $\alpha_1 =0.005$ and $\alpha_2=0.002$.
    The calculation is performed without an ac magnetic field, starting from an initial state slightly tilted from equilibrium.}
\label{fig:LLG_numerical_analysis}
\end{figure}

The results above show that approaching the stability threshold from the stable side enhances the helicity-dependent absorption response. Once the threshold is crossed, however, one of the modes acquires gain and the system can enter a regime of self-oscillation rather than linear response. To examine this regime, we solve the full LLG equation because the linearized treatment is no longer valid when the oscillation amplitude becomes large.
Figure~\ref{fig:LLG_numerical_analysis} shows the real-time magnetization dynamics in the absence of an ac magnetic field for representative parameter sets: (a) in the stable region ($c_J=0$) and (b) in the unstable region ($c_J=0.08 \, {\rm T}$).
In Fig.~\ref{fig:LLG_numerical_analysis}(a), we find that the transverse magnetization components decay to zero.
In contrast, Fig.~\ref{fig:LLG_numerical_analysis}(b) shows that the magnitude of the transverse magnetization components grows and eventually saturates at a value of order unity.

\section{Discussion}
\label{sec:discussion}
\subsection{Role of the EP}

In previously studied gain-loss-type ferromagnetic and synthetic-antiferromagnetic systems, response enhancement is often associated with the exceptional point itself, because eigenmode nonorthogonality and reduced effective damping can develop in the same parameter region. In the present antiferromagnet, by contrast, the resonance-frequency scale is set by the large exchange field \(H_J\). Under the conditions \(\alpha_1,\alpha_2\ll1\) and \(H_J\gg H_0,H_D,c_J\), the damping rate near the exceptional point is approximately
\begin{align}
\Im\omega_{\rm EP}
\simeq
-\frac{\gamma}{2}
\left(
c_J+(\alpha_1+\alpha_2)H_J
\right).
\label{eq:img_omega_EP}
\end{align}
Because the term \((\alpha_1+\alpha_2)H_J\) is amplified by the large exchange field, the effective damping at the exceptional point can remain substantial even when the Gilbert damping itself is small. Thus, unlike in many ferromagnetic non-Hermitian systems, the exceptional point does not necessarily correspond to a weakly damped or strongly amplified response. Instead, the response is governed more directly by approaching the stability threshold, where the imaginary part of one eigenfrequency approaches zero. 
The sign of \(c_J\) selects which helicity branch approaches the stability threshold, so the helicity of the enhanced absorption and the resulting circular polarization can be reversed by the current direction. In this sense, the EP and the stability threshold should be viewed as distinct but related consequences of the same non-Hermitian coupled-mode problem. The EP characterizes the topology of mode coalescence in parameter space, whereas the stability threshold determines where the large, switchable linear response is realized.
This is why the largest absorption and helicity-selective responses in the present system appear mainly near the stability threshold rather than at the exceptional point.
This separation between eigenmode coalescence and damping compensation provides a useful design criterion for non-Hermitian antiferromagnetic resonance: the helicity-selective response is maximized by tuning toward the stability threshold, not necessarily toward the EP.

\subsection{Material considerations}

We have used MnF$_2$ as a representative material because its uniaxial easy-axis anisotropy is consistent with the minimal model considered here. 
However, its N\'eel temperature, \(T_{\rm N}\simeq 67.7\,{\rm K}\), is well below room temperature~\cite{Vaidya2020}. 
From the perspective of device applications, NiO is an attractive alternative because of its much higher N\'eel temperature, \(T_{\rm N}\simeq 523\,{\rm K}\)~\cite{Wang2018NiOAFMR}. 
However, the strong hard-axis anisotropy in NiO confines the N\'eel vector to the easy plane~\cite{Khymyn2017AntiferromagneticTHzFrequencyJosephsonLikeOscillator}, providing a large restoring torque against out-of-plane motion. 
As a result, a substantially larger antidamping SOT is required to destabilize the equilibrium state and reach the stability threshold. 
Another possible candidate is YFeO$_3$, whose N\'eel temperature is approximately \(644\,{\rm K}\)~\cite{Amelin2018YFeO3}. 
However, YFeO$_3$ is a canted antiferromagnet with a finite Dzyaloshinskii--Moriya interaction~\cite{Kim2017YFeO3}, and therefore cannot be described by the minimal collinear model considered here. 
These considerations indicate that MnF$_2$ provides a simple material example for illustrating the mechanism discussed in this work, whereas identifying a room-temperature material compatible with the present model remains an important direction for future study.

\section{Summary}
\label{sec:summary}

We have shown that helicity-selective absorption in a uniaxial antiferromagnetic-insulator/nonmagnetic-metal junction can be realized near the stability threshold. 
When the spin-orbit torque compensates the damping of one helicity mode, the corresponding resonance becomes weakly damped, producing enhanced absorption and linewidth narrowing on the stable side of the threshold. 
Because the sign of the spin-orbit torque selects which helicity mode approaches the threshold, the dominant circular polarization of the transmitted wave can be reversed by reversing the current direction. 
For the MnF$_2$ parameters considered here, the exceptional point is located far from the threshold regime and remains strongly damped, showing that the enhanced response is governed by the stability threshold rather than by EP-induced mode coalescence.
These results establish stability-threshold control as a design principle for helicity-selective antiferromagnetic resonance in uniaxial antiferromagnets, motivating the search for room-temperature easy-axis materials compatible with the present model.

\begin{acknowledgments}
This work was supported by JSPS KAKENHI for Grant Numbers~JP24K06951, JP20H05641, JP21K14551, JP24K01377, JP24H02232, JP24H02235, JP26K01378, and JP24H00400.
\end{acknowledgments}

\appendix

\section{Analytical expression of the stability boundary}
\label{app:analytic}

The critical value of $c_J$ for the stability transition can be calculated analytically, assuming 
$H_D,H_0,|c_J|\ll H_J$ and $\alpha_1,\alpha_2\ll 1$.
We define an average and a difference of the two Gilbert damping constants as $x=(\alpha_1+\alpha_2)/2$ and $y\equiv \alpha_2-\alpha_1$, respectively.
Then, the two Gilbert damping constants are written as
\begin{align}
\alpha_1=x-\frac{y}{2},
\qquad
\alpha_2=x+\frac{y}{2}.
\end{align}
Keeping terms up to the first order in the small quantities \(\alpha_1,\alpha_2\) and \(c_J\), the coefficients of the quadratic equation are approximated as
\begin{align}
a &\simeq 1+iy, \\
b &\simeq 2\gamma H_0 + i\gamma \left[ 2x(H_D+H_J)
+ H_0 y + c_J \right], \\
c &\simeq \gamma^2 Q + i\gamma^2 c_J (H_0 - H_D-H_J),
\end{align}
where
\begin{align}
Q &\equiv (H_0+H_J+H_D)(H_0-H_J-H_D)+H_J^2 \nonumber\\
&= H_0^2-H_D^2-2H_JH_D .
\end{align}
The resonance condition \(F(\omega)=0\) is written as
\begin{align}
F(\omega) &\simeq F_0(\omega)+\delta F(\omega),
\label{eq:F_split} \\
F_0(\omega) &= \omega^2+ 2\gamma H_0 \, \omega+\gamma^2Q, \label{eq:F0_def} \\
\delta F(\omega) &= iy\,\omega^2 +i\gamma\Bigl[2x(H_D+H_J)+H_0y+c_J\Bigr]\omega \nonumber \\
&\qquad+i\gamma^2 c_J (H_0 - H_D- H_J) .
\label{eq:dF_def}
\end{align}
For \(x=y=c_J=0\), the resonant frequencies are obtained from $F_0(\omega)=0$ as
\begin{align}
\omega_\lambda^{(0)} =\gamma \Bigl[-H_0+\lambda \sqrt{H_D(H_D+2H_J)} \Bigr],
\label{eq:omega0_branch_corrected}
\end{align}
where $\lambda = \pm 1$ corresponds to the helicity of the external ac magnetic field that induces the magnetic resonance.
We note that $\omega_\lambda^{(0)}$ is a real number. 
For small values of $x$, $y$, and $c_J$, the solutions are modified as $\omega_\lambda =
\omega_\lambda^{(0)}+\delta\omega_\lambda$, where the imaginary part of $\delta\omega_\lambda$ is given by
\begin{align}
\Im \delta \omega_\lambda &= -\frac{\gamma}{2\lambda R} \Bigl[ (\lambda R-H_0) \left\{2(H_D+H_J)x+\lambda R y \right\} \nonumber\\
&\hspace{10mm} + (\lambda R-H_D-H_J)c_J \Bigr],
\label{eq:Imomega_threshold}
\end{align}
where $R = \sqrt{H_D(H_D+2H_J)}$.
The critical value of $c_J$ for the stability transition of the helicity mode $\lambda$ is determined from $\Im\omega_{\lambda}=0$ as
\begin{align}
c_{J,{\rm cr}}^{(\lambda)} = - \frac{(\lambda R-H_0) \left[ 2(H_D+H_J)x+\lambda R y \right]}{\lambda R-H_D-H_J}.
\label{eq:cJ_threshold}
\end{align}
The stability boundary becomes a straight line in the $(\alpha_1,c_J)$ diagram for a fixed $\alpha_2$, because $c_{J,{\rm cr}}^{(\lambda)}$ is a linear function of $x$ and $y$, both of which are linear combinations of $\alpha_1$ and $\alpha_2$.

\section{Location of the EP}
\label{app:EP}

\begin{figure}[t]
  \centering
  \includegraphics[width=1.0\linewidth]{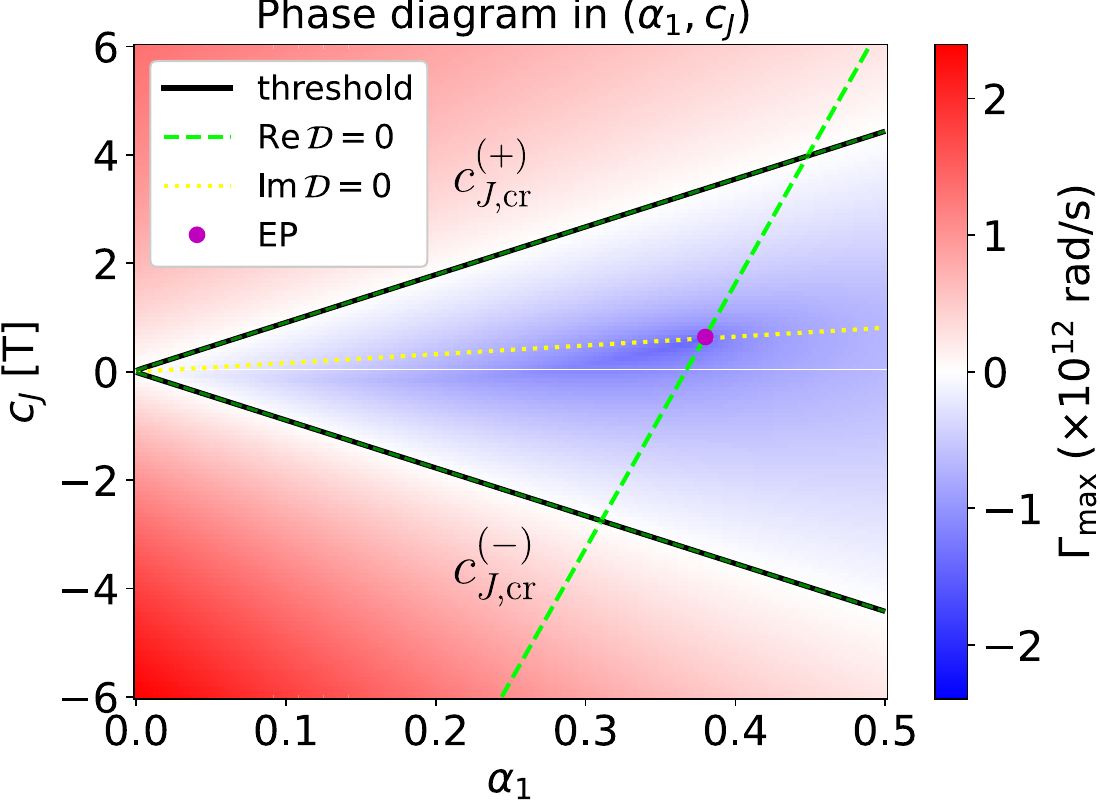}
  \caption{
  Stability diagram in the $(\alpha_1,c_J)$ plane
for $H_0=0$ and $\alpha_2=0.002$ in the extended region.
The color map shows
$\Gamma_{\max}=\max[\mathrm{Im}\,\omega_+,\mathrm{Im}\,\omega_-]$.
The black lines indicate the stability thresholds
$\Gamma_{\max}=0$.
The dashed and dotted lines represent
$\mathrm{Re}\,\mathcal{D}=0$ and
$\mathrm{Im}\,\mathcal{D}=0$, respectively.
The solid magenta circle at their intersection corresponds to the exceptional point (EP).
}
  \label{fig:phase_diagram_EP}
\end{figure}

The EP is defined by the discriminant condition $\mathcal D = b^2-4ac =0$.
To estimate the EP location analytically, we define
\begin{align}
    x=\frac{\alpha_1+\alpha_2}{2},
    \qquad
    y=\alpha_2-\alpha_1.
\end{align}
Under the conditions
$H_D,H_0,|c_J|\ll H_J$ and $\alpha_1,\alpha_2\ll1$,
the EP condition gives
\begin{align}
x_{\mathrm{EP}}
&\simeq
\sqrt{\frac{2H_D}{H_J}}
+
\frac{H_0-H_D}{2H_DH_J}c_J,
\label{EP_condition_x}
\\
y_{\mathrm{EP}}
&\simeq
-\frac{H_0}{H_D}
\sqrt{\frac{2H_D}{H_J}}
-\left[
\frac{1}{2H_D}
+
\frac{H_0(H_0-H_D)}
{2H_D^2H_J}
\right]c_J.
\label{EP_condition_y}
\end{align}
For the MnF$_2$ parameters
$H_J=47.05~\mathrm{T}$ and $H_D=0.82~\mathrm{T}$,
with $H_0=0$ and $\alpha_2=0.002$,
Eqs.~(\ref{EP_condition_x}) and (\ref{EP_condition_y}) give $\alpha_{1,\mathrm{EP}}\simeq0.359$ and $c_{J,\mathrm{EP}}\simeq0.585~\mathrm{T}$. Since $\alpha_{1,\mathrm{EP}}\simeq0.359$ is not small,
this estimate lies outside the strict validity of the small-damping approximation.
Solving $\mathcal D = b^2-4ac =0$ directly gives
$\alpha_{1,\mathrm{EP}}\simeq0.379$ and
$c_{J,\mathrm{EP}}\simeq0.613~\mathrm{T}$.

Figure~\ref{fig:phase_diagram_EP} shows the extended
$(\alpha_1,c_J)$ diagram, where the green dashed line and the yellow dotted line indicate 
\({\rm Re}\,{\cal D}=0\) and \({\rm Im}\,{\cal D}=0\), respectively. 
The EP lies far outside the parameter range shown in Fig.~\ref{fig:phase_diagram},
\(\alpha_1\leq0.01\) and \(|c_J|\leq0.08~\mathrm{T}\).
At the EP, \(\mathrm{Im}\,\omega_{\mathrm{EP}}\simeq-1.45\times10^{12}~\mathrm{rad/s}\), indicating that the modes near the EP are strongly damped. 
Therefore, the EP is not expected to produce the sharp helicity-selective absorption enhancement discussed in the main text.

\bibliography{ref}
\end{document}